\documentclass
[twocolumn,superscriptaddress,showpacs,preprintnumbers,amsmath,amssymb]{revtex4}

\usepackage{graphicx}
\usepackage{dcolumn}
\usepackage{bm}

\begin{document}

\preprint{APS/123-QED}

\title{Pressure-induced changes in the magnetic and valence state of EuFe$_{2}$As$_{2}$}

\author{K.~\textsc{Matsubayashi}}
\affiliation{Institute for Solid State Physics, The University of Tokyo, Kashiwanoha, Kashiwa, Chiba 277-8581, Japan}
\affiliation{JST, TRIP, 5 Sanbancho, Chiyoda, Tokyo 102-0075, Japan}
 \author{K.~\textsc{Munakata}}
\affiliation{Institute for Solid State Physics, The University of Tokyo, Kashiwanoha, Kashiwa, Chiba 277-8581, Japan}
\affiliation{JST, TRIP, 5 Sanbancho, Chiyoda, Tokyo 102-0075, Japan}
\author{N.~\textsc{Katayama}} \altaffiliation[Present address:]{Department of Physics, University of Virginia}
 \affiliation{Institute for Solid State Physics, The University of Tokyo, Kashiwanoha, Kashiwa, Chiba 277-8581, Japan}
\author{M.~\textsc{Isobe}}
\affiliation{Institute for Solid State Physics, The University of Tokyo, Kashiwanoha, Kashiwa, Chiba 277-8581, Japan}
\affiliation{JST, TRIP, 5 Sanbancho, Chiyoda, Tokyo 102-0075, Japan}
\author{K.~\textsc{Ohgushi}}
\affiliation{Institute for Solid State Physics, The University of Tokyo, Kashiwanoha, Kashiwa, Chiba 277-8581, Japan}
\affiliation{JST, TRIP, 5 Sanbancho, Chiyoda, Tokyo 102-0075, Japan}
\author{Y.~\textsc{Ueda}}
\affiliation{Institute for Solid State Physics, The University of Tokyo, Kashiwanoha, Kashiwa, Chiba 277-8581, Japan}
\affiliation{JST, TRIP, 5 Sanbancho, Chiyoda, Tokyo 102-0075, Japan}
\author{N.~\textsc{Kawamura}}
\affiliation{Japan Synchrotron Radiation Research Institute (JASRI/SPring-8), 1-1-1 Kouto, Sayo, Hyogo 679-5198, Japan}
\author{M.~\textsc{Mizumaki}}
\affiliation{Japan Synchrotron Radiation Research Institute (JASRI/SPring-8), 1-1-1 Kouto, Sayo, Hyogo 679-5198, Japan}
\author{N.~\textsc{Ishimatsu}}
\affiliation{Department of Physical Science, Graduate School of Science, Hiroshima University, 1-3-1 Kagamiyama, Higashi-Hiroshima, Hiroshima 739-8526, Japan}
\author{ M. \textsc{Hedo}}
\affiliation{Faculty of science, University of the Ryukyus, Nishihara, Okinawa 903-0213, Japan}
\author{I.~\textsc{Umehara}}
\affiliation{Department of Physics, Faculty of Engineering, Yokohama National University, Yokohama 240-85013, Japan}
\author{Y.~\textsc{Uwatoko}}
\affiliation{Institute for Solid State Physics, The University of Tokyo, Kashiwanoha, Kashiwa, Chiba 277-8581, Japan}
\affiliation{JST, TRIP, 5 Sanbancho, Chiyoda, Tokyo 102-0075, Japan}


\begin{abstract}
We present the results of electrical resistivity, ac specific heat, magnetic susceptibility, X-ray absorption spectroscopy (XAS) and X-ray magnetic circular dichroism (XMCD) of the ternary iron arsenide EuFe$_{2}$As$_{2}$ single crystal under pressure. Applying pressure leads to a continuous suppression of the antiferromagnetism associated with Fe moments and the antiferromagnetic transition temperature becomes zero in the vicinity of a critical pressure $P_{\rm c}$~$\sim$~2.5-2.7 GPa. Pressure-induced re-entrant superconductivity, which is highly sensitive to the homogeneity of the pressure, only appears in the narrow pressure region in the vicinity of $P_{\rm c}$ due to the competition between superconductivity and the antiferromagnetic ordering of Eu$^{2+}$ moments. The antiferromagnetic state of Eu$^{2+}$ moments changes to the ferromagnetic state above 6~GPa. We also found that the ferromagnetic order is suppressed with further increasing pressure, which is connected with a valence change of Eu ions.
\end{abstract}

\maketitle
\section{Introduction}
Since the discovery of superconductivity in LaFeAsO$_{1-x}$F$_{x}$~\cite{Kamihara}, a large number of new superconductors with FeAs layers have been reported. In the oxygen-free iron arsenide $A$Fe$_{2}$As$_{2}$ ($A$ = Ca, Sr, Ba, Eu) with the tetragonal ThCr$_{2}$Si$_{2}$ structure, the antiferromagnetism of Fe moments is suppressed by chemical doping or applying pressure, and superconductivity appears above a critical concentration and pressure~\cite{Rotter2,Alireza}. 
Among the  $A$Fe$_{2}$As$_{2}$ compounds, EuFe$_{2}$As$_{2}$ is an interesting system from the viewpoint of magnetism; Eu ion is in a divalent state with magnetic moments and orders antiferromagnetically below 20~K~\cite{Jeevan1}. In this system, superconductivity emerges in 50~$\%$ K-doped sample at $T_{\rm c}$ $\sim$ 32~K while short range magnetic ordering of the Eu moments coexists with the superconducting state below 15~K~\cite{Jeevan2,Anupam}. Hence, EuFe$_{2}$As$_{2}$ offers an unique opportunity to study an interplay between the superconductivity and magnetism of both Eu and Fe ions.

	According to the high pressure experiment on parent EuFe$_{2}$As$_{2}$ compound, the resistivity drop at $\sim$29~K suggesting the onset of superconductivity is observed at pressures above~2.0 GPa~\cite{Miclea}. With further decreasing temperature, the resistivity increases again and exhibiting a maximum, interpreted to re-entrant superconductivity caused by the magnetic ordering of the the Eu$^{2+}$ moments as observed in the rare-earth nickel borocarbides~\cite{Eisaki}. More recently, zero resistivity and full shielding effect in the ac magnetic susceptibility have been observed at a critical pressure $P_{\rm c}$ $\sim$ 2.5~GPa \cite{Terashima}. The discrepancy for the appearance of zero resistivity is probably due to the difference in high pressure experimental conditions. In fact, in the case of $A$Fe$_{2}$As$_{2}$ ($A$ = Ca, Sr, Ba), there exists a crucial difference in the presence/absence of superconductivity depending on the hydrostaticity of pressure~\cite{Yu, Kotegawa, Matsubayashi, Yamazaki}.
Therefore, high pressure study under highly hydrostatic conditions is demanded on EuFe$_{2}$As$_{2}$ to obtain the intrinsic properties. In addition, the maximum pressure in the above previous reports on EuFe$_{2}$As$_{2}$ was limited below 3~GPa. As for the competition between the superconductivity and magnetism of Eu ions, it would be also rewarding to extend the pressure range to higher pressure because sufficiently high pressure induces the valence change from Eu$^{2+}$ (4$f^7$, magnetic) to Eu$^{3+}$ (4$f^6$, non-magnetic) state, which may be more favorable for the appearance of superconductivity. According to high pressure experiment for Eu-metal, pressure-induced superconductivity appears at sufficiently high pressure where Eu valence state might be in a trivalent or mixed-valent state \cite{Debessai}. In order to clarify these points and construct the phase diagram in the wider pressure range, we have measured resistivity, magnetic susceptibility, ac specific heat, X-ray absorption spectroscopy (XAS) and X-ray magnetic circular dichroism (XMCD) up to 23~GPa on a single crystal of EuFe$_{2}$As$_{2}$ grown by the self-flux method.

\section{EXPERIMENTAL}
Single crystals were grown by an FeAs self-flux method. The starting materials were put into an alumina crucible with the ratio Eu:Fe:As = 1:5:5 and sealed in a double quartz tube. The tube was heated up to 1120$^\circ$C, and slowly cooled down to 900$^\circ$C in 72 hours.

	High pressure was generated by using two types of pressure cells: a hybrid piston cylinder clamped cell and a cubic anvil cell \cite{Uwatoko, Mori}. The pressure at low temperature in the piston cylinder cell was determined by the pressure dependence of a superconducting transition temperature of Sn~\cite{Jennings}. The applied pressure inside the cubic anvil cell is calibrated by the measurement of the resistivity changes of Bi and Te associated with their structural phase transitions at room temperature. In order to maintain a constant pressure, the force applied to the sample is kept constant during the measurement by cooling and warming runs. We note that the reason why we used two kinds of pressure cell is to check the effect of the inhomogeneity of pressure on EuFe$_{2}$As$_{2}$. The piston cylinder clamped cell with a liquid pressure medium is commonly used to apply hydrostatic pressure, however, the solidification of the liquid pressure-transmitting medium causes inhomogeneous pressure distributions and uniaxial stress because of the uniaxial geometry of the pressure cell. On the other hand, the cubic anvil apparatus is known to generate hydrostatic pressure owing to the multiple anvil geometry; a gasket with a Teflon cell, in which the sample is immersed in the liquid pressure-medium, is compressed from three directions with six tungsten carbide anvils \cite{Mori}. The importance of the hydrostatic pressure is demonstrated by the pressure-induced superconductivity EuFe$_{2}$As$_{2}$ (see below for details).

	Electrical resistivity was measured by a standard four-probe technique with current flow in the $ab$-plane. The dc magnetization was measured using a commercial (MPMS) magnetometer. The ac magnetic susceptibility was measured at a fixed frequency of 307~Hz with a modulation field of 0.3~mT applied along $ab$-plane. Ac calorimetric measurement is performed by Joule heating type technique, and experimental details are described in elsewhere \cite{Matsubayashi2}. In the above high pressure experiments, pressure transmitting mediums for the piston cylinder cell and the cubic anvil cell were Daphne7373 and glycerin, respectively.
	
	X-ray absorption (XAS) and X-ray magnetic circular dichroism (XMCD) measurements at the Eu $L_{3}$-edge were performed under pressure at the beamline BL39XU of SPring-8, Japan \cite{Kawamura}. Single crystalline sample was cut into 70~$\times$~50~$\times$~10~$\mu$m$^{3}$ for the transmission measurement. The sample was loaded in a diamond anvil cell (DAC), filled with glycerin served as a pressure transmitting medium. Pressure calibration was performed using the fluorescence from ruby chips mounted with the sample inside the DAC. X-ray beam and magnetic field was aligned parallel to the $c$-axis.
		
\section{RESULTS AND DISCUSSION}
\subsection{Electrical resistivity and specific heat}

\begin{figure}[t]
\begin{center}
\includegraphics[width=0.5\textwidth]{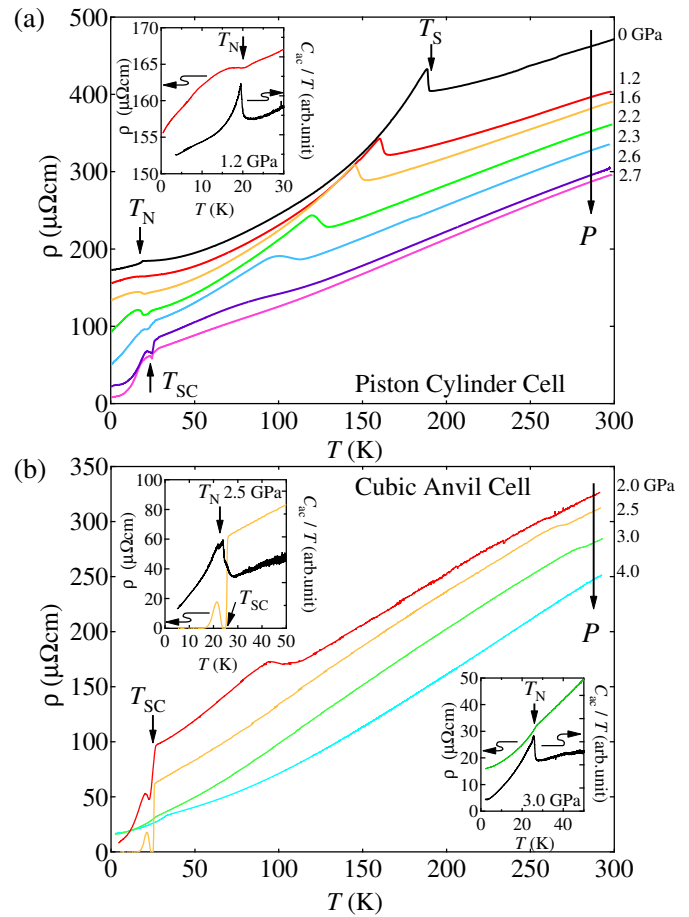}
\end{center}
\caption{(Color online) Temperature dependence of the electrical resistivity $\rho$ for EuFe$_{2}$As$_{2}$ under pressure using (a) piston cylinder cell and (b) cubic anvil cell. The arrows at $T_{\rm s}$ and $T_{\rm N}$ indicate the location of the structural/antiferromagnetic transition temperature of Fe moments and antiferromagnetic transition temperature of Eu moments, respectively. The insets show the temperature dependence of $\rho$ and $C_{ac}/T$ at selected pressures.}
\label{}
\end{figure}

\begin{figure}[t]
\begin{center}
\includegraphics[width=0.5\textwidth]{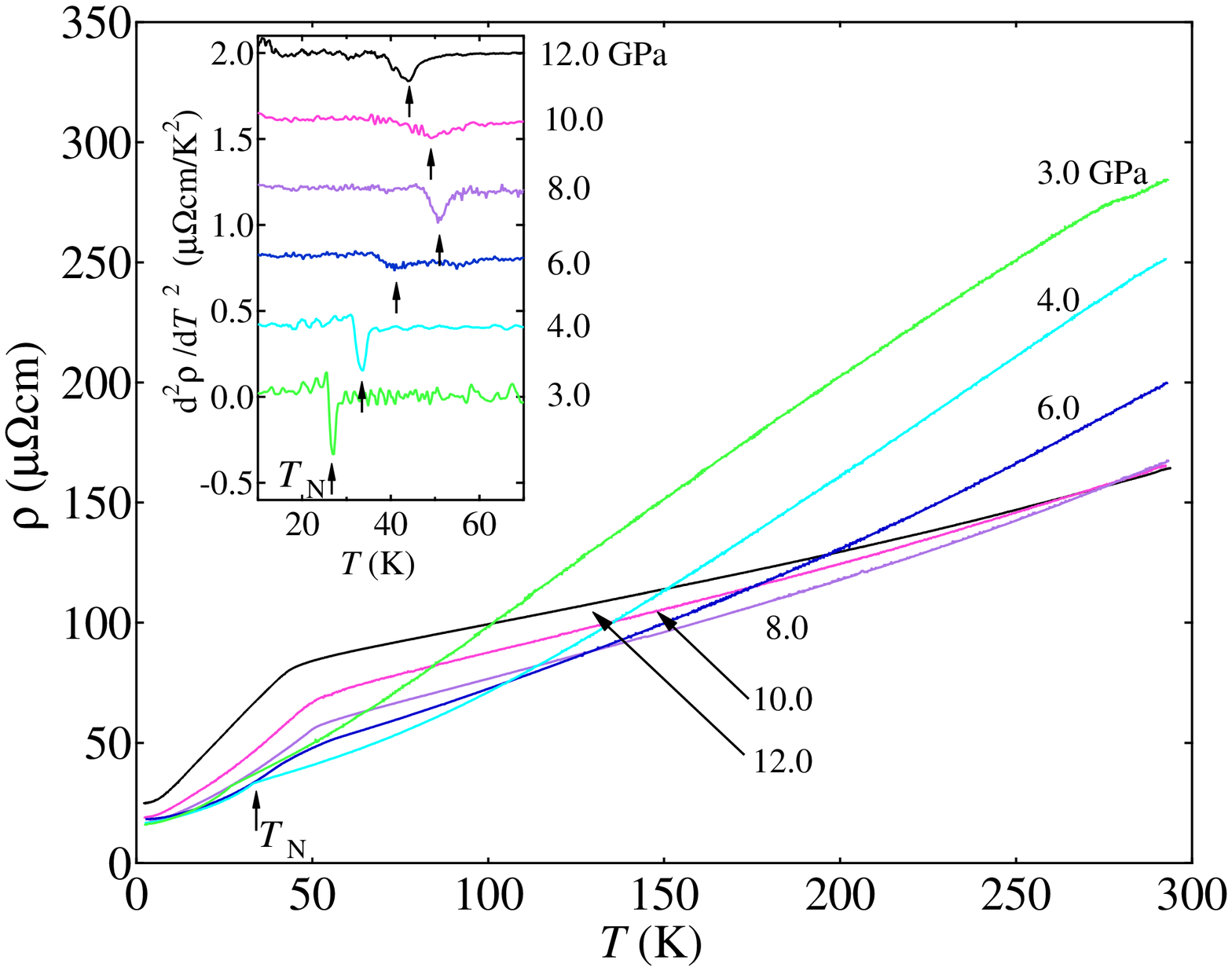}
\end{center}
\caption{(Color online) Temperature dependence of $\rho$ above 3 GPa using cubic anvil cell. The inset shows $d^{2}\rho(T)$/$dT^{2}$ versus $T$ below 70 K. The curves for different pressures are shifted for clarity.}
\label{}
\end{figure}

Figure 1 shows the temperature dependences of the electrical resistivity $\rho(T)$ for EuFe$_{2}$As$_{2}$ at high pressure using (a) piston cylinder cell and (b) cubic anvil cell. At ambient pressure, $\rho(T)$ shows step-like increase at $T_{\rm s}$~$\sim$~185~K, where the structural transition and itinerant antiferromagnetism of Fe moments occur simultaneously \cite{Jeevan1}. The sharpness of the transition indicates the high quality of our sample. With increasing pressure, $T_{\rm s}$ shifts to lower temperature and seems to disappear at a critical pressure $P_{\rm c}$ $\sim$ 2.5-2.7~GPa in both high pressure experiments. The broad transition feature in the vicinity of $P_{\rm c}$ caused by combined effect of a first-order nature of this phase transition and the inhomogeneity of the pressure distribution. At ambient pressure, a clear kink observed at $T_{\rm N}$~$\sim$~19~K indicates the antiferromagnetic transition of Eu$^{2+}$ moments (see Fig.~1 (a)). For pressures higher than 1.2~GPa, $\rho(T)$ shows a distinct increase at around 20 K, reaching a maximum and decreasing again. Here we assign this sharp dip to $T_{\rm N}$ so as to be consistent with specific heat as shown in the inset of Fig.~1~(a). Above 2~GPa, a new anomaly appears at $T_{\rm SC}$~$\sim$~25~K, implying the appearance of the superconductivity as reported previously. The resistivity drop is more pronounced at 2.6~GPa, however, the resistance does not reach to the zero resistance. By contrast, in the case of using cubic anvil cell, the superconducting transition with zero resistance is achieved at almost the same pressure and temperature region as shown in Fig.~1~(b). At lower temperatures, the resistivity has a peak at around 20~K, coinciding with the specific heat anomaly due to the antiferromagntic ordering of Eu$^{2+}$ moments. Therefore, the reentrant behavior of the resistivity is caused by the competition of the superconductivity and the antiferromagnetic state of Eu$^{2+}$ moments. With further increasing pressure above 3~GPa, the superconducting transition becomes no longer visible, and then resistivity and specific heat anomaly corresponding to the AF ordering of Eu$^{2+}$ moments survives (see the lower inset of Fig.~1~(b)).

	Here we compare our results with the previous high pressure studies. The resistivity measurement made by Miclea $et~al.$ agrees with our result using piston cylinder cell, although the zero resistance was not achieved in both cases. It is noteworthy that the pressure transmitting medium used above these experiments solidifies below $P_{\rm c}$($\sim$2.5-2.7~GPa); the solidification pressure of silicone oil (Miclea $et~al.$) and Daphne7373 (our experiment using piston cylinder) is 1.0~GPa and 2.2~GPa, respectively \cite{Sandberg, Murata}. It is important to keep the pressure medium in a liquid state to generate hydrostatic pressure because the shear stress of a solid medium causes nonhydrostatic pressure. Actually, the zero-resistivity was detected by Terashima $et~al.$ using Daphne7474, which remains liquid up to 3.7 GPa \cite{Murata}. This is consistent with our results using cubic anvil cell. Note that we use glycerin served as a pressure medium, which is in a liquid state up to $\sim$5~GPa \cite{Drozd}. From these results, we conclude that the pressure homogeneity gives a critical influence on the detection of the superconductivity as a consequence of the occurrence of the superconductivity in a narrow pressure window between 2.5 and 3.0~GPa.

	Here we recall that the high pressure studies on CaFe$_{2}$As$_{2}$; superconductivity appears only under non-hydrostatic pressure conditions \cite{Yu}, contrary to our observation in EuFe$_{2}$As$_{2}$. In the case of CaFe$_{2}$As$_{2}$, a collapsed tetragonal ($cT$) structure, resulting in significant anisotropic change in the lattice constant, is identified above $P_{\rm c}$. The superconductivity observed under nonhydrostatic pressure originates in a low temperature multicrystallographic-phase state that includes not only the nonsuperconducting $cT$ phase but also the superconducting tetragonal phase \cite{Torikachvili}. On the other hand, recent the x-ray diffraction measurement under pressure in EuFe$_{2}$As$_{2}$ revealed that $cT$ phase is identified above 8~GPa \cite{Uhoya}. Therefore, the $cT$ phase is absent in the vicinity of $P_{\rm c}$ in EuFe$_{2}$As$_{2}$, suggesting the different origin of highly sensitive to pressure inhomogeneity between EuFe$_{2}$As$_{2}$ and CaFe$_{2}$As$_{2}$. From these results, the pressure-induced superconductivity in EuFe$_{2}$As$_{2}$ is stabilized only in the narrow pressure region due to the competition between superconductivity and the magnetic order of Eu ions. Therefore, a precise tuning of pressure without pressure inhomogeneity is needed to observe the superconductivity.

To further elucidate the nature of the high pressure region, we have performed resistivity measurements using cubic anvil cell up to 12~GPa. For pressures lower than 4~GPa, $\rho(T)$ follows simple metallic behavior at high temperatures followed by a resistivity drop at $T_{\rm N}$. Here, we define $T_{\rm N}$ as the minimum in the second derivative $d^{2}\rho(T)$/$dT^{2}$ (see the inset of Fig.~2). $T_{\rm N}$ shifts toward higher temperatures on application of pressure reaching $\sim$50 K at 8 GPa, and then decreases slightly. Note that the overall shape of $\rho(T)$ changes at higher pressures, which is related to the instability of the Eu valence as described in the following.
	
\subsection{Magnetic susceptibility}

\begin{figure}[t]
\begin{center}
\includegraphics[width=0.45\textwidth]{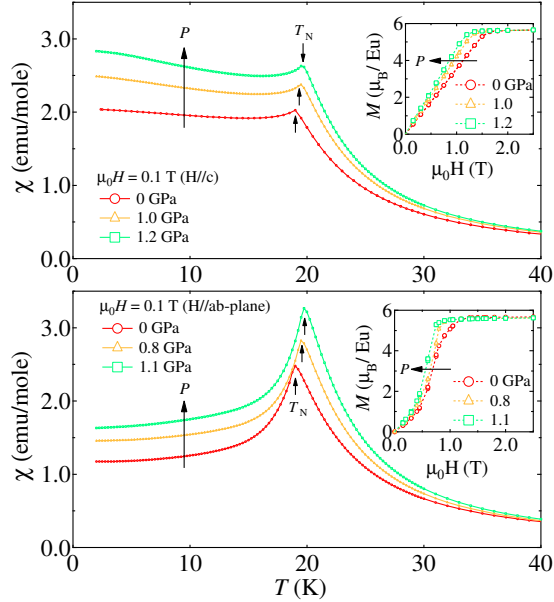}
\end{center}
\caption{(Color online) Temperature dependence of dc magnetic susceptibility under magnetic field (a) $H$~$\bot$~$c$ and (b) $H$~$\|$~$ab$-plane at selected pressure. The inset shows $M$-$H$ curves at 2~K.
}
\label{}
\end{figure}

Figure 3 shows the temperature dependence of magnetic susceptibility $\chi(T)$ of EuFe$_{2}$As$_{2}$ under pressure for an applied magnetic field of 0.1~T along perpendicular and parallel to $ab$-plane. At ambient pressure, the high temperature $\chi(T)$ is isotropic and follows the Curie-Weiss law, $\chi(T)$~=~$\chi_{\rm 0}$~+~$C$/($T$-$\theta_{\rm W}$), where $\chi_{\rm 0}$ is the temperature-independent susceptibility, $C$ is the Curie-Weiss constant and $\theta_{\rm W}$ is the Curie-Weiss temperature. The effective moments are close to the value of a free divalent Eu ion, and $\theta_{\rm W}$ $\sim$ 16~K, as reported previously. A pronounced kink at $T_{\rm N}$~$\sim$~19 K in both orientations indicates the antiferromagnetic transition of Eu$^{2+}$ moments. $T_{\rm N}$ slightly increases with increasing pressure, which is consistent with the results of aforementioned resistivity and specific heat. As shown in the inset of Fig.~3, a metamagnetic transition is found for both field orientations at temperatures below $T_{\rm N}$ \cite{Jiang,Tokunaga}. Here we note that no hysteresis was found within the experimental precision. According to the neutron scattering experiment, the magnetic structure of the antiferromagnetic order is $A$-type; Eu$^{2+}$ spins align ferromagnetically in the basal planes while the planes are coupled antiferromagnetically \cite{Xiao}. The critical field of the metamagnetic transition decreases with increasing pressure, suggesting that the interlayer antiferromagnetic coupling becomes weaker under pressure.

	To clarify the nature of the magnetic properties at higher pressures, we have carried out ac magnetic susceptibility measurement under pressure. Figure~4 shows the temperature dependence of the real $\chi_{\rm ac}^{'}$ and imaginary part $\chi_{\rm ac}^{''}$ of the ac magnetic susceptibility at constant pressure. At 2.0~GPa, $\chi_{\rm ac}^{'}$ exhibits a kink at a temperature $T_{\rm N}$. At $P_{\rm c}$($\sim$2.5~GPa), where the SC transition appears in the resistivity, three distinct anomalies are present as shown in the inset of Fig. 4. $\chi_{\rm ac}^{'}$ shows a hump at $\sim$25~K well corresponding to zero resistance. A kink appears at $T_{\rm N}$~$\sim$~21~K close to the temperature where the maximum in the resistivity and the specific heat anomaly, reflecting the occurrence of the antiferromagnetic order. A large diamagnetic signal due to the SC transition is observed at lower temperature, which corresponds to nearly 100~$\%$ superconducting shielding by comparing to the diamagnetic signal of Pb with almost the same size as the sample. With further increasing pressure above 3.0~GPa, the SC diamagnetic signal is no longer visible, and $T_{\rm N}$ increases with increasing pressure, consistent with the aforementioned resistivity and specific heat measurements. At 6.0~GPa a rather peculiar behavior is observed; the kink at $T_{\rm N}$ becomes less pronounced, and we found a peak in $\chi_{\rm ac}^{''}$ at $\sim$30~K. At higher pressure of 8.0 GPa a pronounced maxima is observed in both $\chi_{\rm ac}^{'}$ and $\chi_{\rm ac}^{''}$ at $T_{\rm Curie}$~$\sim$~48~K, indicating the occurrence of the ferromagnetic order.  It is worth comparing our results to substituted system, such as EuFe$_{2-x}$Ni$_{x}$As$_{2}$ and EuFe$_{2}$(As$_{0.7}$P$_{0.3}$)$_{2}$, in which the magnetic ordering of Eu$^{2+}$ moments evolves from antiferromangtic to ferromagnetic state above a critical concentration \cite{Ren1,Ren2}. 

\begin{figure}[t]
\begin{center}
\includegraphics[width=0.45\textwidth]{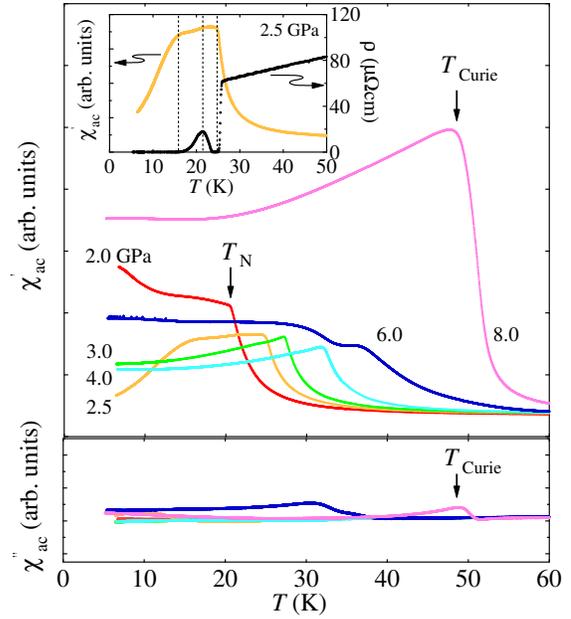}
\end{center}
\caption{(Color online) Temperature dependence of the real $\chi_{\rm ac}^{'}$ and the imaginary $\chi_{\rm ac}^{''}$ parts of the ac magnetic susceptibility for EuFe$_{2}$As$_{2}$ under pressure. }
\label{}
\end{figure}

\subsection{XAS and XMCD}
Figure~5 shows Eu~$L_{3}$-edge XAS spectra of EuFe$_{2}$As$_{2}$ at 15 and 300~K under various pressures. At around 2~GPa and 5~GPa, the XAS spectrum exhibits a single peak structure corresponding to Eu$^{2+}$ ions, which is consistent with the divalent nature of Eu reported at ambient pressure \cite{Anupam}. On the other hand, the spectra for pressures above $\sim$9~GPa consist of two resolved peaks, indicating the mixed valence character of the Eu ions. The relative intensity of the trivalent state absorption peak to that of the divalent state increases with increasing pressure. The averaged valence was determined by fitting the XAS spectra to an arctangent step function and a Lorentzian peak for each valence state. As shown in the middle panel of Fig.~5, Eu valence monotonically increases above $\sim$9~GPa whereas there is no significant valence change at lower pressures. We note that the valence of the Eu ions exhibits the temperature dependence at $\sim$15~GPa. This is consistent with the decrease of $T_{\rm Curie}$ above 10~GPa, indicating a valence fluctuating Eu state.

\begin{figure}[t]
\begin{center}
\includegraphics[width=0.5\textwidth]{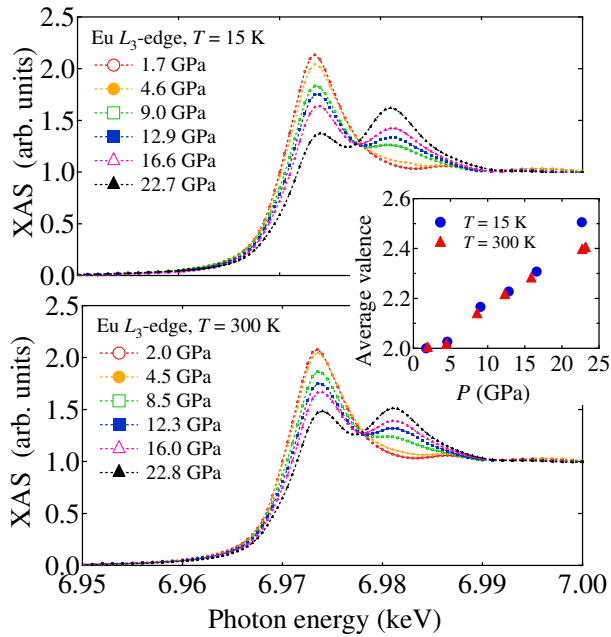}
\end{center}
\caption{(Color online) XAS spectra at various pressures at 15 and 300~K for the Eu~$L_{3}$-edge. The middle-panel shows the pressure dependence of the averaged valence. Dotted lines are a guide to eye.}
\label{}
\end{figure}

\begin{figure}[t]
\begin{center}
\includegraphics[width=0.5\textwidth]{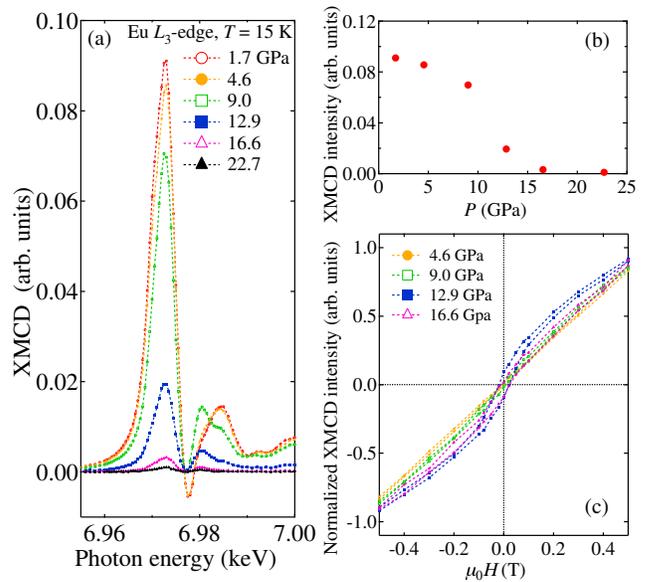}
\end{center}
\caption{(Color online) (a) XMCD spectra at various pressures at 15~K for the Eu~$L_{3}$-edge. (b) Pressure dependence of XMCD intensity at 6.973~keV. (c) Normalized XMCD intensity is plotted as functions of the magnetic field at selected pressures.}
\label{}
\end{figure}

	In the left panel of Fig.~6, we show the XMCD spectra at 15~K in magnetic fields of 0.6~T. The main peak at around 6.973~keV is attributed to Eu$^{2+}$ state, while the higher energy peak is complicated by the presence of neighboring atoms. Actually, it is known from both experimental and theoretical studies that the XMCD spectrum is affected by the hybridization between the Eu 5$d$ and Fe 3$d$ bands \cite{Asakura}. A detailed theoretical study is needed to understand the mechanism of the XMCD. Hereafter, we consider the main peak at 6.973~keV. The observation of XMCD signal indicates the presence of magnetic moment of the absorbing magnetic atoms, and XMCD signal is proportional to the projection of the magnetization vector along the x-ray wave vector. Therefore, XMCD signal is basically absent in antiferromagnetic state. However, the applied magnetic field may induce canting of the AFM spin structure \cite{Jiang}, as observed in our magnetization measurements, and results in the observation of XMCD signal at low pressures. Above 9~GPa, the amplitude of the XMCD peak obviously decreases and the overall shape of the XMCD spectra at higher energy (6.98-6.99~keV) is modified. These observations can be interpreted as the reduction of the effective magnetic moments of Eu as a result of a dilution of the magnetic Eu$^{2+}$ with increasing pressure. As shown in Fig.~6 (b), with further increasing pressure the XMCD intensity seems to disappear around $\sim$20~GPa, suggesting the collapse of the magnetism of Eu. In order to obtain the information on the magnetic state, we performed the element-specific magnetization (ESM) measurements at 6.973~keV, which is corresponding to the peak of XMCD signal due to the Eu$^{2+}$ state. As shown in Fig.~6 (c), XMCD intensity increases almost linearly at 4.6~GPa, on the other hand, at higher pressures a small hysteresis loop is observed and appears to be approaching saturation. This is consistent with the ferromagnetic state as detected by the ac susceptibility measurement.

\subsection{Pressure-temperature phase diagram}
\begin{figure}[t]
\begin{center}
\includegraphics[width=0.45\textwidth]{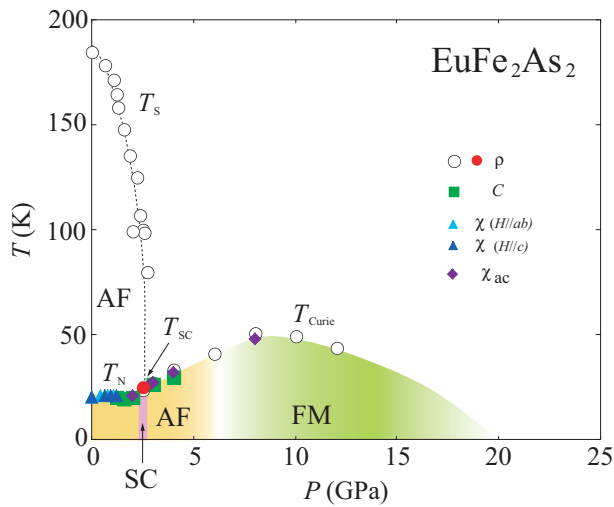}
\end{center}
\caption{(Color online) Pressure-temperature phase diagram of EuFe$_{2}$As$_{2}$ derived from the electrical resistivity (open and closed circles), ac specific heat (squares), and magnetic susceptibility (triangles and diamonds). Dotted lines are a guide to eye.}
\label{Phase diagram}
\end{figure}
	The results on EuFe$_{2}$As$_{2}$ single crystals studied here are summarized in Fig.~7 as a pressure-temperature phase diagram. The antiferromagnetism associated with Fe moments is suppressed at $P_{\rm c}$~$\sim$~2.5-2.7~GPa while the antiferromagnetic order of localized Eu moments is almost pressure independent below $P_{\rm c}$. Our key finding is that $T_{\rm SC}$ only appears in the narrow pressure region in the vicinity of $P_{\rm c}$, indicating that the SC state competes with the AF state of Eu$^{2+}$ moments. Above 3~GPa, $T_{\rm N}$ of Eu$^{2+}$ moments starts to increase, and a pressure-induced transition from antiferromagnetic to ferromagnetic ordering is confirmed by the ac magnetic susceptibility and ESM measurements. The magnetic coupling between the Eu moments would be mediated by the indirect Ruderman-Kittel-Kasuya-Yosida (RKKY) exchange interaction in EuFe$_{2}$As$_{2}$. It is conceivable that the RKKY exchange interaction could be affected by the distance between Eu moments with applying pressure, resulting in the pressure-induced variation from antiferromagnetic to ferromagnetic order due to the sign change of the RKKY interaction. It is found that the ferromagnetic transition temperature $T_{\rm Curie}$ shows a maximum at around 8~GPa, and decreases at higher pressures. Combining the results of XAS and XMCD data, the suppression of the ferromagnetic state is connected with the valence change from magnetic Eu$^{2+}$ to nonmagnetic Eu$^{3+}$ state. According to the recent the x-ray diffraction measurement under pressure in EuFe$_{2}$As$_{2}$, collapsed tetragonal ($cT$) phase is found above 8~GPa \cite{Uhoya}. It is known that the pressure-induced structural transition toward the $cT$ phase is connected with the valence change of Eu ions, as reported in EuFe$_{2}$P$_{2}$ and EuCo$_{2}$P$_{2}$ \cite{Ni}. Therefore, we conclude that pressure-induced valence change occurs at $\sim$8~GPa. To obtain further insight in this critical pressure region, high pressure bulk measurements will be needed to complete the phase diagram.
	
\section{SUMMARY}
	In summary, we performed the electrical resistivity, ac specific heat, magnetic susceptibility, XAS and XMCD of EuFe$_{2}$As$_{2}$ single crystal under pressure. The antiferromagnetism associated with Fe moments collapses at the critical pressure $P_{\rm c}$~$\sim$~2.5-2.7~GPa, and pressure-induced re-entrant superconductivity only appears in the narrow pressure region in the vicinity of $P_{\rm c}$. Consequently, we conclude that the difference of the pressure homogeneity has a critical influence on the appearance of superconductivity in EuFe$_{2}$As$_{2}$. The antiferromagnetic order of Eu$^{2+}$ moments changes to ferromagnetic order above around 6~GPa, confirmed by ac magnetic susceptibility and element selective magnetization (ESM) measurements using XMCD. We found that the ferromagnetic order is suppressed with further increasing pressure connected with a valence change of Eu ions.

\section*{ACKNOWLEDGMENTS}
We thank H. S. Suzuki, G. Cao and J. S. Schilling for helpful discussions and comments. We also thank S. Suzuki for experimental assistance. This work was partially supported by the approval of the Japan Synchrotron Radiation Research Institute (JASRI) (Proposal No. 2009B1959), and a Grant-in-Aid for Research (No. 21340092, No. 20102007, No. 19GS0205) from the Ministry of Education, Culture, Sports, Science and Technology, Japan.

\end{document}